\documentclass[prd,preprint,superscriptaddress,amsmath,amssymb,nofootinbib]{revtex4}
\pdfoutput=1

\usepackage{graphicx,color,slashed}

\def\be{\begin{eqnarray}}
\def\ed{\end{eqnarray}}

\begin{document}


\title{\bf \Large Slight excess at 130 GeV in search for a charged Higgs boson decaying to a 
charm quark and a bottom quark at the Large Hadron Collider}

\author{A.G. Akeroyd}
\email{a.g.akeroyd@soton.ac.uk}
\affiliation{School of Physics and Astronomy, University of Southampton,
Highfield, Southampton SO17 1BJ, United Kingdom}

\author{Stefano Moretti}
\email{S.Moretti@soton.ac.uk; stefano.moretti@physics.uu.se}
\affiliation{School of Physics and Astronomy, University of Southampton,
Highfield, Southampton SO17 1BJ, United Kingdom}
\affiliation{Department of Physics and Astronomy, Uppsala University, Box 516, SE-751 20 Uppsala, Sweden}

\author{Muyuan Song}
\email{2106393210@pku.edu.cn}
\affiliation{Center for High Energy Physics,
Peking University, Beijing 100871, China}
\affiliation{School of Physics and Astronomy, University of Southampton,
Highfield, Southampton SO17 1BJ, United Kingdom}

\date{\today}

\begin{abstract}
\noindent
Searches for a charged Higgs boson ($H^\pm$) decaying to a charm quark and a bottom quark ($H^\pm \to cb$)
have been carried out at the Large Hadron Collider (LHC) in the decay of top quarks ($t\to H^\pm b$). In a recent search by
the ATLAS collaboration (with all Run II data, 139 fb$^{-1}$) a local excess of around $3\sigma$ has been observed, 
which is best fitted by a charged Higgs boson with a mass ($m_{H^\pm}$) of around 130 GeV and a product of branching ratios (BRs) given by
BR$(t\to H^\pm b)\times{\rm BR}(H^\pm\to cb)=0.16\%\pm 0.06\%$. In the context of Two-Higgs-Doublet Models  (2HDM) with independent
Yukawa couplings for $H^\pm$
we present the parameter space for which this excess (assuming it to be genuine) can be accommodated, taking into account the limits from LHC searches for
$H^\pm\to cs$ and $H^\pm\to \tau\nu$ at $m_{H^\pm}$=130 GeV and the constraint from $b\to s\gamma$.
It is then shown that such an excess cannot be explained in 2HDMs with natural flavour conservation, but
can be accommodated in the flipped Three-Higgs-Doublet Model (3HDM) and in the aligned 2HDM (A2HDM). Upcoming searches
with 139 fb$^{-1}$ in the channels $H^\pm\to cb$ (CMS), $H^\pm \to cs$ (ATLAS/CMS) and $H^\pm \to \tau\nu$ (ATLAS/CMS)
will determine if the excess is the first sign of an $H^\pm$ with $m_{H^\pm}=130$ GeV.
\end{abstract}

\maketitle

\section{Introduction}
\noindent
In the year 2012 the discovery of a new particle with a mass of around 125 GeV was announced by the ATLAS and CMS collaborations of the Large Hadron Collider (LHC)
\cite{Aad:2012tfa,Chatrchyan:2012xdj}. Ongoing (Run I and Run II) measurements of its properties are in very good agreement 
(within experimental error) with those of the Higgs boson of the 
Standard Model (SM). In particular, it has been established that the spin of the 125 GeV particle is zero (and hence it is a boson), and 
five decay channels ($\gamma\gamma$, $ZZ$, $WW$, $\tau\tau$, and $bb$) 
have now been observed with a statistical significance of greater than 
$5\sigma$ (e.g. see \cite{ATLAS:2018kot}). 
The above branching ratios (BRs) are in good agreement with 
those of the SM Higgs boson, although the current experimental precision
allows for small deviations from these BR predictions in the SM.
In addition, the four main
production mechanisms (gluon-gluon fusion, vector boson $(W/Z)$ 
fusion, associated production with a vector boson, and 
associated production with top quarks) have been measured, 
with no significant deviation so far from the predicted cross-sections of the SM Higgs boson.
Measurements of all the above BRs and cross-sections with the full Run II data (139 fb$^{-1}$ at $\sqrt s=13$ TeV) have been combined
to show a signal strength relative to that of the SM Higgs boson of $1.02^{+0.07}_{-0.06}$ \cite{CMS:RunII} (CMS) and $1.06\pm 0.06$ \cite{ATLAS:RunII} (ATLAS).

If the observed 125 GeV boson is indeed the 
(solitary) Higgs boson of the SM then the ongoing (and future)
experimental measurements of its properties would converge to the precise theoretical predictions
for this particle. However, it is possible that 
the 125 GeV boson is the first scalar to be discovered from a non-minimal Higgs 
sector i.e. the scalar potential contains additional scalar isospin doublets 
or higher representations such as scalar isospin triplets. 
In this scenario, future measurements (e.g. at the High Luminosity LHC and/or at a future $e^+e^-$ collider) of the cross sections and BRs of the 125 GeV boson
could start to show increasingly significant deviations from those
of the SM Higgs boson. Moreover, there would also be the possibility of 
discovering additional neutral scalars, or physical charged scalars ($H^\pm$) that are present in such
enlarged Higgs sectors. 

In this work we shall focus on the searches for an $H^\pm$ from models with additional isospin doublets.
In the context of a Two-Higgs-Doublet Model (2HDM) 
the lack of observation of 
an $H^\pm$ at the LHC rules out parameter space of $\tan\beta$ 
(which is present in the Yukawa couplings) and $m_{H^\pm}$, where $\tan\beta=v_2/v_1$, 
and $v_1$ and $v_2$ are the vacuum expectation values (VEVs) of the two 
Higgs doublets respectively (for reviews see e.g. 
\cite{Branco:2011iw,Akeroyd:2016ymd}). 
The mass of an $H^\pm$ could be above or below the mass of the top quark ($m_t$), and searches in both
these scenarios have been carried out at the LHC. In the former case
a production mechanism for $H^\pm$ would be via $t\to H^\pm b$, and it is on this process that we focus.
In a 2HDM with Natural Flavour Conservation (NFC) \cite{Glashow:1976nt} one expects
the dominant decay channels of an $H^\pm$ with $m_{H^\pm}\le m_t$ to be $H^\pm\to cs$ and $H^\pm\to \tau\nu$.
Although the partial decay width for the decay $H^\pm \to cb$ has an enhancement factor relative to the above channels due to $m_b>m_\tau,m_c,m_s$, this decay channel has suppression from the small 
Cabibbo-Kobayashi-Maskawa (CKM) matrix 
element $V_{cb}$ $(|V_{cb}|<<|V_{cs}|)$.
In three of the four types of 2HDM with NFC, it has been known for a long time that
BR($H^\pm\to cb$) is small (of the order of $1\%$)  \cite{Barger}.
In the flipped 2HDM (which is one of the four types) a large BR($H^\pm\to cb$) would be possible for large $\tan\beta$ and $m_{H^\pm} < m_t$ 
(this was first explicitly pointed out in  \cite{Grossman,Akeroyd:1994ga}), but
the strong constraint $m_{H^\pm}> 500$ GeV from $b\to s\gamma$ \cite{Ciuchini1,Ciuchini2,Borzumati,Gambino,Misiak,Misiak2} ensures that $H^\pm\to tb$ is open and hence is the dominant decay due to
 $|V_{tb}|\approx 1$ and the large value of $m_t$.

In 2HDMs with NFC the interactions of $H^\pm$ with fermions at tree-level are determined by one unknown parameter $\tan\beta$ and 
the phenomenology of $H^\pm$ will also depend on $m_{H^\pm}$. As first pointed out in \cite{Grossman} and subsequently developed in previous works by some of us \cite{Akeroyd:1994ga,Akeroyd:1995cf,Akeroyd:1998dt,Akeroyd2,Akeroyd:2016ssd,Akeroyd:2018axd},
it is possible for $H^\pm\to cb$ to become the dominant decay channel for $m_{H^\pm}\le m_t$ in models with more than two Higgs doublets (while still keeping NFC) and also comply with the constraint from $b\to s\gamma$.
The same result is true in the aligned 2HDM \cite{Pich:2009sp} (A2HDM, which does not have NFC but suppresses flavour changing neutral currents by a different
mechanism). 
This difference in phenomenology of $H^\pm$ is because the Yukawa couplings in the latter two models depend on more than one parameter 
i.e. in a Three-Higgs-Doublet Model (3HDM) with NFC there
are four parameters that determine the Yukawa couplings of $H^\pm$, while in the A2HDM there are five parameters.

The first search for $H^\pm\to cb$ decays originating from $t\to H^\pm b$ was carried out by the CMS collaboration in
\cite{CMS:2018dzl} with 20 fb$^{-1}$ of data at $\sqrt s=8$ TeV; limits on BR($t\to H^\pm b)\times$BR($H^\pm\to cb$)
in the range $0.3\%$ to $1.4\%$ were obtained (with a dependence on $m_{H^\pm}$). 
 Recently the ATLAS collaboration
has carried out such a search with 139 fb$^{-1}$ of data at $\sqrt s=13$ TeV \cite{ATLAS:2021zyv}, and obtained limits
on BR($t\to H^\pm b)\times {\rm BR}(H^\pm\to cb$)
in the range $0.15\%$ to $0.42\%$. Of interest is a local excess of $3\sigma$ significance ($1.6\sigma$ global) which is best
fitted by $m_{H^\pm}=130$ GeV and  BR($t\to H^\pm b)\times{\rm BR}(H^\pm\to cb)=0.16\%\pm 0.06\%$. In our previous works \cite{Akeroyd2,Akeroyd:2016ssd,Akeroyd:2018axd}
the magnitude of BR($t\to H^\pm b)\times{\rm BR}(H^\pm\to cb)$ as a function of the parameters that determine the Yukawa couplings in 
2HDMs and 3HDMs was studied, and the parameter space which would be probed in future LHC searches for $t\to H^\pm b$ with $H^\pm\to cb$ was depicted. Building on the results of these works
and assuming that the above excess is genuine, in this work we quantify the parameter space 
which gives the above best-fit value for the product of BRs with $m_{H^\pm}=130$ GeV.


This work is organised as follows. 
In section II we give an introduction to the phenomenology 
of the lightest $H^\pm$ in 2HDMs/3HDMs with NFC and in the A2HDM.
In section III the searches for $t\to H^\pm b$ 
at the LHC are summarised, with attention given to the search for $H^\pm \to cb$ and the
$3\sigma$ excess at  $m_{H^\pm}=130$ GeV.
In section IV our results are presented, and conclusions are contained in
section V.

\section{Parameter space for a large BR$(H^\pm\to cb)$ in 2HDMs and 3HDMs}
\noindent
In this section the parameter space for a large BR$(H^\pm\to cb)$ is identified in the 
models under consideration (2HDMs and 3HDMs).
In section IIA, the fermionic couplings of $H^\pm$ are discussed. These couplings depend on the masses of the fermions in the interaction,
the relevant CKM matrix element, and the parameters of the scalar potential. In section IIB, the constraints on these fermionic couplings of $H^\pm$ are summarised
in each model. In section IIC, explicit formulae for the BRs of 
the decay of $H^\pm$ to fermions are given, and the condition for a large BR$(H^\pm\to cb)$ is described. The discussion in this section is an updated version of
equivalent discussions from our earlier works in \cite{Akeroyd2,Akeroyd:2016ssd,Akeroyd:2018axd}. A detailed review of the 2HDM is presented in \cite{Branco:2011iw}, and an increasing number of works
are now focussing on various aspects of 3HDMs (both theoretical and phenomenological) e.g. see \cite{Logan,Ivanov:2010wz}, with recent studies in  \cite{Das:2019yad,Gomez-Bock:2021uyu,Jurciukonis:2021wny,Chakraborti:2021bpy,Ivanov:2021pnr,Buskin:2021eig,Darvishi:2021txa,Das:2021oik,Boto:2021qgu,Khater:2021wcx,Kalinowski:2021lvw}.

\subsection{Fermionic couplings of $H^\pm$ in the 2HDMs and 3HDMs}
\noindent
The Lagrangian in a 2HDM and in a 3HDM that describes the interactions of $H^\pm$ with 
the fermions (the Yukawa couplings) can be written as follows (e.g. \cite{Grossman}):
\begin{equation}
{\cal L}_{H^\pm} =
-\left\{\frac{\sqrt2V_{ud}}{v}\overline{u}
\left(m_d X{P}_R+m_u Y{P}_L\right)d\,H^+
+\frac{\sqrt2m_\ell }{v} Z\overline{\nu_L^{}}\ell_R^{}H^+
+{H.c.}\right\}\,.
\label{lagrangian}
\end{equation}
Here $u(d)$ refers to the up(down)-type quarks (i.e. $u$ refers to the up, charm and top quark and similar for $d$),  and $\ell$
refers to the electron, muon and tau. Other symbols represent  i) chirality projection operators ($P_L$ and $P_R$); ii) CKM matrix element ($V_{ud}$); iii) the VEV of the Higgs boson
in the SM ($v$=246 GeV). The parameters $X$, $Y$ and $Z$ contain the dependence on the parameters of the scalar potential.
In a 2HDM there is only one $H^\pm$, while in a 3HDM there are two $H^\pm$s which are usually
labelled as $H^\pm_1$ and $H^\pm_2$ (with $m_{H^\pm_1}< m_{H^\pm_2}$)\footnote{Sometimes these are labelled as $H^\pm_2$ and $H^\pm_3$ e.g.  \cite{Logan}.}. In the latter case, eq.~(\ref{lagrangian}) would be 
modified to have $X_i$, $Y_i$, and $Z_i$ for each $H^\pm_i$.
Flavour changing neutral currents (FCNCs) that are mediated by scalars at tree-level must be strongly suppressed in order to comply with
experiment. Such neutral currents can be eliminated
by requiring that the Yukawa couplings are invariant under certain discrete symmetries (NFC, mentioned earlier), 
a framework in which each fermion type receives its mass from one VEV only \cite{Glashow:1976nt}.
The charge assignments of the scalar and
fermion fields under the discrete symmetries can be found in many works e.g.\cite{Akeroyd:2016ssd}.
The requirement of NFC (which is not the only way to suppress FCNCs to an acceptable level - see A2HDM \cite{Pich:2009sp} to be discussed later) leads to four distinct 2HDMs \cite{Barger}: Type I, Type II, lepton-specific, and
flipped. In Table~\ref{couplings} the couplings $X$, $Y$, and $Z$ in these 
2HDMs are given, and each coupling depends on just one parameter ($\tan\beta$) of the scalar potential.
\begin{table}[h]
\begin{center}
\begin{tabular}{|c||c|c|c|}
\hline
& $X$ &  $Y$ &  $Z$ \\ \hline
Type I
&  $-\cot\beta$ & $\cot\beta$ & $-\cot\beta$ \\
Type II
& $\tan\beta$ & $\cot\beta$ & $\tan\beta$ \\
Lepton-specific
& $-\cot\beta$ & $\cot\beta$ & $\tan\beta$ \\
Flipped
& $\tan\beta$ & $\cot\beta$ & $-\cot\beta$ \\
\hline
\end{tabular}
\end{center}
\caption{The couplings $X$, $Y$, and $Z$ in the Yukawa interactions of $H^\pm$ in the four versions of the 2HDM with NFC.}
\label{couplings}
\end{table}
In contrast, the couplings $X$, $Y$, and $Z$ of
the lightest $H^\pm$ in a 3HDM with NFC are functions of four parameters of the scalar potential.
This can be understood as follows.
A unitary matrix $U$ connects the charged scalar fields in the 
weak eigenbasis ($\phi^\pm_1,\phi^\pm_2,\phi^\pm_3)$ with the 
physical scalar fields
($H^\pm_1$, $H^\pm_2$) and the charged Goldstone boson $G^\pm$ as follows:

\begin{equation}
	\left( \begin{array}{c} G^+ \\ H_1^+ \\ H_2^+ \end{array} \right) 
	= U \left( \begin{array}{c} \phi_1^+ \\ \phi_2^+ \\ \phi_3^+ \end{array} \right).
	\label{eq:Udef}
\end{equation}
The couplings of $H^\pm_1$ are as follows \cite{Logan}: 
\begin{equation}
	X = \frac{U_{d2}^\dagger}{U_{d1}^\dagger}, \quad \quad 
	Y = - \frac{U_{u2}^\dagger}{U_{u1}^\dagger}, \quad \quad 
	Z = \frac{U_{\ell 2}^\dagger}{U_{\ell 1}^\dagger}\,.
\label{eq:xyz}
\end{equation}
The values of $d$, $u$, and $\ell$ in eq.~(\ref{eq:xyz}) (not to be confused with the notation for the quarks and leptons themselves in eq.~(\ref{lagrangian})) in these matrix elements  of $U^\dagger$ are listed in Table \ref{valuesudl} and
depend on which of the five distinct 3HDMs with NFC  is under consideration e.g.
\begin{table}[h]
\begin{center}
\begin{tabular}{|c||c|c|c|}
\hline
& $u$ &  $d$ &  $\ell$ \\ \hline
3HDM (Type I) &  2 & 2 & 2 \\
3HDM (Type II) & 2 & 1 & 1 \\
3HDM (Lepton-specific) & 2 & 2 & 1 \\
3HDM (Flipped) & 2 & 1 & 2 \\
3HDM (Democratic) & 2 & 1 & 3 \\
\hline
\end{tabular}
\end{center}
\caption{The five versions of the 3HDM with NFC,
and the corresponding values of $u$, $d$, and $\ell$. The choice of $u=2$ means that the up-type quarks receive their mass
from the vacuum expectation value $v_2$, and likewise for $d$ (down-type quarks) and $\ell$ (charged leptons).}
\label{valuesudl}
\end{table}
the choice of $d=1$, $u=2$, and $\ell=3$ means that the down-type quarks receive their mass from the vacuum expectation value $v_1$,
the up-type quarks from $v_2$, and the charged leptons from $v_3$ (a choice called the ``democratic 3HDM''). The other
possible choices of $d$, $u$, and $\ell$ in a 3HDM with NFC are given the same names as the four types of 2HDM.
The couplings of the $H^\pm_2$ (i.e. the heavier charged scalar) are obtained from eq.~(\ref{eq:xyz}) by
making the replacement $2 \to 3$ in the numerators of $X$, $Y$, and $Z$. 
The matrix $U$ can be written explicitly as a function of
four parameters  $\tan\beta$, $\tan\gamma$, $\theta$, and $\delta$, where
\begin{equation}
	\tan\beta = v_2/v_1, \qquad \tan\gamma = \sqrt{v_1^2 + v_2^2}/v_{3}\,.
\end{equation}
and $v_1$, $v_2$, and $v_3$ are the VEVs of the three scalar doublets.
The angle $\theta$ and the complex phase $\delta$ can be expressed explicitly as  
functions of several parameters in the scalar potential \cite{Logan}.
The explicit form of $U$ is:
\begin{eqnarray}
	U &=& \left( \begin{array}{ccc} 
		1 & 0 & 0 \\
		0 & e^{-i \delta} & 0 \\
		0 & 0 & 1 \end{array} \right)
		\left( \begin{array}{ccc}
		1 & 0 & 0 \\
		0 & c_\theta & s_\theta e^{i \delta} \\
		0 & -s_\theta e^{-i \delta} & c_\theta \end{array} \right)
		\left( \begin{array}{ccc}
		s_\gamma & 0 & c_\gamma \\
		0 & 1 & 0 \\
		-c_\gamma & 0 & s_\gamma \end{array} \right)
		\left( \begin{array}{ccc}
		c_\beta & s_\beta & 0 \\
		-s_\beta & c_\beta & 0 \\
		0 & 0 & 1 \end{array} \right)
	\nonumber \\
	&=& \left( \begin{array}{ccc}
	s_\gamma c_\beta & s_\gamma s_\beta 	& c_\gamma \\
	-c_\theta s_\beta e^{-i\delta} - s_\theta c_\gamma c_\beta 
		& c_\theta c_\beta e^{-i\delta} - s_\theta c_\gamma s_\beta & s_\theta s_\gamma \\
	s_\theta s_\beta e^{-i\delta} - c_\theta c_\gamma c_\beta 
		& -s_\theta c_\beta e^{-i\delta} - c_\theta c_\gamma s_\beta & c_\theta s_\gamma 
	\end{array} \right).
	\label{eq:Uexplicit}
\end{eqnarray}
Here $s$ and $c$ denote the sine or cosine of the respective angle (e.g. $s_\beta$ is $\sin\beta$). Hence the functional forms of the couplings
$X$, $Y$, and $Z$ in a 3HDM with NFC depend on four parameters. As mentioned earlier, in 
a 2HDM with NFC these couplings only depend on $\tan\beta$ due to the analogous matrix $U$ now being a $2\times 2$ matrix with elements 
that depend on $\sin\beta$ and $\cos\beta$ only.

The A2HDM is a 2HDM in which NFC is not imposed \cite{Pich:2009sp}. Instead, both scalar doublets ($\Phi_1$ and $\Phi_2$)
couple to all types of fermions, but tree-level FCNCs are eliminated due to an
alignment of the Yukawa couplings of  $\Phi_1$ and $\Phi_2$. The interaction of $H^\pm$ with the fermions
in the A2HDM is also described by eq.~(\ref{lagrangian}).
However, the couplings $X$, $Y$, and $Z$ in the A2HDM are
determined by five independent parameters instead of the four parameters of the 3HDM. Moreover, in contrast to the 3HDM, these five parameters do not arise from a unitary matrix $U$ and thus 
they are not constrained by the requirement $UU^\dagger=I$, as pointed out in \cite{Logan}.  In the A2HDM, the magnitudes $|X|$, $|Y|$, and $|Z|$ may be taken as
independent input parameters. 
We shall be presenting results for three classes of models: i) the case of a 2HDM with independent couplings $X$, $Y$ and $Z$, an example being
the A2HDM;
ii) the 2HDM with NFC; ii) the 3HDM with NFC.

\subsection{Constraints on the couplings $X$, $Y$, and $Z$}
\noindent
The couplings $X$, $Y$, and $Z$ (and combinations thereof) are constrained
from various processes. We summarise here the bounds (which are also summarised in \cite{Logan,Akeroyd2,Akeroyd:2016ssd,Akeroyd:2018axd}) that will be used when generating our results in section IV.

In the context of the A2HDM (for which  $|X|$, $|Y|$, and $|Z|$ are independent parameters) a detailed study can be found in Refs.~\cite{Jung:2010ik,Trott:2010iz}. 
To a good approximation, these constraints can be directly applied to the lightest $H^\pm_1$ of a 3HDM provided that the 
contribution from the heavier $H^\pm_2$ to a given process is considerably 
smaller (e.g. if $m_{H^\pm_2}\gg m_{H^\pm_1}$). We will comment below on the inclusion of $H^\pm_2$ on the constraints obtained from $b\to s\gamma$ and
the electric dipole moment of the neutron.

The strongest constraint on the coupling $Y$ is from the measurement of the process $Z\to b\overline b$
at the LEP experiment. For $m_{H^\pm}<m_t$ (i.e. the scenario on which we focus)
the constraint is roughly $|Y|<0.8$ (with the assumption that $|X|\le 50$, which ensures that
the dominant contribution to $Z\to b\overline b$ is from the $Y$ coupling). 
The coupling $X$ is also constrained from $Z\to b\overline b$ to be roughly $|X|\le 50\to 100$ for $m_{H^\pm}<m_t$ (with a dependence on $|Y|$). However,
more important to this work are the constraints on the plane $[X,Y]$ from $t\to H^\pm b$ (for which $|Z|$ also has an influence due its effect on the BRs of $H^\pm$), as shown in 
\cite{Akeroyd2,Akeroyd:2018axd} and in section IV. A recent study of the contribution of the scalars (both neutral and charged) in a 3HDM to $Z\to b\overline b$ has been carried out in \cite{Jurciukonis:2021wny}.

The measured BR of the rare decay $b\to s\gamma$ provides a constraint 
on the combination Re$(XY^*)$ 
\begin{equation}
-1.1 \le {\rm Re}(XY^*)\le 0.7.
\label{bsgamma}
\end{equation}
This constraint was derived in \cite{Trott:2010iz} for  
$m_{H^\pm}=100$ GeV and will be slightly weaker for $m_{H^\pm}=130$ GeV. It is an approximation for the case when 
i) the contribution from $|Y|^2$ can be neglected (which
is a fairly good approximation because $|Y|<0.8$ for $m_{H^\pm}< m_t$ as mentioned above) and ii)
Im($XY^*$) is small (which is also a good approximation due to the electric dipole moment of the neutron, as shown below shortly).
Constraints on the $H^\pm$ contribution
to $b\to s \gamma$ in the A2HDM without these two
approximations are studied in \cite{Jung:2010ik}. 
Other works on the effect of $H^\pm$ on $b\to s \gamma$ are usually in the 
context of the 2HDM with NFC (early works in  \cite{Grinstein:1987pu,Grinstein:1987vj,Hou:1987kf,Barger,Bertolini:1990if})
and include various higher-order
corrections to both the SM contribution and the $H^\pm$ contribution \cite{Ciuchini1,Ciuchini2,Borzumati,Gambino,Misiak,Misiak2}.

The electric dipole moment of the neutron (a CP violating observable) provides the following constraint
on Im$(XY^*)$ \cite{Trott:2010iz}:
\begin{equation} 
|{\rm Im}(XY^*)|\le 0.1\,.
\label{Imxy}
\end{equation}
This bound is for $m_{H^\pm}=100$ GeV and is an order-of-magnitude
estimate. Other constraints (such as $|Z|\le 40$ and $|XZ|\le 1080$ from processes involving leptons \cite{Logan}) have very little impact on our study.
In our numerical analysis in section IV we will respect all the above constraints. In the majority of our results the couplings $X$, $Y$, $Z$ will be taken to be real, and so
the constraint from the  electric dipole moment of the neutron will be automatically satisfied.

In a 3HDM one would have contributions to 
$b\to s\gamma$ and the electric dipole moment of the neutron from both $H^\pm_1$ and $H^\pm_2$. 
As mentioned earlier, if $m_{H^\pm_2}\gg m_{H^\pm_1}$ then it is a good approximation to apply the above constraints on $X$, $Y$ and $Z$ to
$H^\pm_1$ alone. Studies of BR($b\to s\gamma$) in 3HDMs including the contribution from both $H^\pm_1$ and $H^\pm_2$
have been carried out to next-to-leading order accuracy
in our previous work in \cite{Akeroyd:2016ssd,Akeroyd:2020nfj}, and these results will be used
 in our study of the 3HDM. The contribution to the electric dipole moment of the neutron from both $H^\pm_1$ and $H^\pm_2$ was studied in
our previous work in \cite{Logan:2020mdz}, and is relevant when the couplings $X_i$ and $Y_i$ $(i=1,2)$ have an imaginary part.

\subsection{The Branching Ratios of $H^\pm$}
\noindent
Only the decays of $H^\pm$ to two fermions will be considered in this work.
In a 2HDM or 3HDM there exist additional neutral CP-even scalars and CP-odd scalars and one or more of these could be lighter than an $H^\pm$ of mass 130 GeV e.g. 
 a CP-odd $A^0$, the presence of which could give rise to the decay $H^\pm\to A^0W^{(*)}$ with a magnitude that is
 determined by the mass splitting of $m_{H^\pm}$ and $m_{A^0}$ (the coupling $H^\pm A^0 W$ is a constant gauge coupling).
 The discovered 125 GeV boson is a CP-even scalar, and if it is the lightest CP-even scalar in a 2HDM or 3HDM then this is called the "normal" scenario.
 If the 125 GeV boson is not the lightest CP-even scalar and is instead one of the heavier CP-even scalars (labelled by $H^0$, called the "inverted scenario") then 
the decay channel $H^\pm \to h^0W^*$ with $m_{h^0}<125$ GeV ($h^0$ being the lightest CP-even scalar that has not been discovered yet)
would be open. In 2HDMs the $H^\pm h^0W$ coupling is proportional to $\cos(\beta-\alpha)$ and so it would be maximised for the inverted scenario
of (the discovered) $H^0$ having SM-like couplings i.e. $\cos(\beta-\alpha)\approx 1$ with  $m_{H^0}=125$ GeV. 
Studies of the case where $H^\pm \to h^0W^*$ and/or $H^\pm\to A^0W^*$ have a  
sizeable (or even dominant) BR can be found in
\cite{Akeroyd:1998dt,Moretti:1994ds,Djouadi:1995gv,Akeroyd2,Kling:2015uba,Arhrib:2016wpw,Arbey:2017gmh,Arhrib:2017wmo,Bahl:2021str,Cheung:2022ndq}.
We assume that these decays are negligible/absent, and this is achieved by i) taking $m_{A^0}>m_{H^\pm}$ in both the normal and inverted scenarios and ii) 
in the inverted scenario by having a small
mass splitting $m_{H^\pm} - m_{h^0}$  which would suppress $H^\pm \to h^0W^*$. If either (or both) of these
decays has a sizeable BR then the BRs of $H^\pm$ to two fermions would be decreased significantly. 
The decay $t\to H^\pm b$ with $H^\pm \to A^0W$ (i.e. an on-shell $W$) has been searched for at the LHC for the decay mode
$A^0\to \mu^+\mu^-$ \cite{CMS:2019idx}. No searches at the LHC have yet been carried out for the decay modes $A^0\to \tau^+\tau^-$ and $A^0\to b\overline b$ from $t\to H^\pm b$, although
such searches were carried out at the Tevatron and LEP2 respectively.
The decay $A^0\to b\overline b$ from $t\to H^\pm b$ with $H^\pm \to A^0W$ would perhaps contribute at some level to the signal for 
 $t\to H^\pm b$ with $H^\pm \to cb$, as mentioned in \cite{Akeroyd2}.

 We note that the decay $H^\pm\to h^0 W^*/H^0 W^*$ to the (discovered CP-even) 125 GeV boson (either $h^0$ or $H^0$, depending on whether one has normal or inverted scenario)  is open for our case of interest of $m_{H^\pm}=130$ GeV. However,
its partial width is very small due to the strong phase space suppression from the small mass splitting of around 5 GeV. Moreover, the couplings $H^\pm h^0 W$ and  $H^\pm H^0 W$ would both be suppressed
as the $h^0$ or $H^0$ are both SM-like in this case, and thus the trigonometric factors in these couplings would both be very close to zero.

 In a 2HDM/3HDM the tree-level expressions for the partial widths of the decay modes of $H^\pm$ to fermions which are lighter than the top quark
(and so the phase space suppression factor can be neglected) are given by (e.g. see \cite{Djouadi:1995gv,Branco:2011iw,Choi:2021nql}):
\begin{equation}
\Gamma(H^\pm\to \ell^\pm\nu)=\frac{G_F m_{H^\pm} m^2_\ell |Z|^2}{4\pi\sqrt 2}\;, 
\label{width_tau}
\end{equation}
\begin{equation}
\Gamma(H^\pm\to ud)=\frac{3G_F V_{ud}m_{H^\pm}(m_d^2|X|^2+m_u^2|Y|^2)}{4\pi\sqrt 2}\;.
\label{width_ud}
\end{equation}

In the expression for 
$\Gamma(H^\pm\to ud)$ the running quark masses are evaluated at the 
scale ($Q$) of $m_{H^\pm}$, and this encompasses the bulk of the QCD corrections.
There are also QCD vertex corrections which multiply the partial widths by
$(1+17\alpha_s/(3\pi))$. 
A study of the BRs as a function of $|X|$, $|Y|$, and $|Z|$ 
was first given in \cite{Akeroyd:1994ga} and more recently in \cite{Akeroyd2,Akeroyd:2018axd}.
For $|X|\gg |Y|,|Z|$ the decay channel BR$(H^\pm\to cb)$ dominates
(which was first mentioned in \cite{Grossman}, although no numerical study was carried out), reaching
a maximum of around $80\%$. In this limit (in which $|X|$ is the dominant coupling), it can be easily shown from
eq.~(\ref{width_ud}) that the ratio of the partial widths of
$H^\pm\to cb$ and $H^\pm\to cs$ is given by
\begin{equation}
\frac{{\rm BR}(H^\pm\to cb)}{{\rm BR}(H^\pm\to cs)}\sim \frac{|V_{cb}|^2 m^2_b}{|V_{cs}|^2m_s^2}\;.
\end{equation}
The value of $m_s$ evaluated at the scale of $m_{H^\pm}$ is crucial in determining the 
magnitude of BR$(H^\pm \to cb)$, with smaller values of $m_s(Q=m_{H^\pm})$ giving rise to a larger BR$(H^\pm \to cb)$.
The importance of the value of $m_s$ in determining the BRs of $H^\pm$ is a unique feature (in phenomenology of Higgs bosons in general)
of this specific scenario of $|X|>>|Y|, |Z|$. The world averages of lattice calculations \cite{Aoki:2021kgd} of $m_s$ give
$m_s(Q=2 \,{\rm GeV})=92.2\pm 1.0\,{\rm MeV}$ for $N_f=2+1$ ($N_f$ is number of flavours) and  
$m_s(Q=2 \,{\rm GeV})=93.40\pm 0.57\,{\rm MeV}$ for $N_f=2+1+1$. Taking $m_s(Q=2 \,{\rm GeV})=93\;{\rm MeV}$ one finds that 
$m_s(Q=130\,{\rm GeV})\approx 55\,{\rm MeV}$, leading to BR$(H^\pm\to cb)\approx 80\%$ (with $m_b(Q=130\,{\rm GeV})= 2.95\, {\rm GeV})$. 
In a 2HDM with NFC the only model (of the four) 
which contains a parameter space for a large BR$(H^\pm\to cb)$ with $m_{H^\pm}< m_t$
is the flipped model. This possibility was mentioned in \cite{Grossman,Akeroyd:1994ga,Akeroyd:1998dt}
and studied in more detail in \cite{Aoki:2009ha,Logan:2010ag} (but using a larger $m_s$ than the above value of $55$ MeV).
However, in the flipped 2HDM the $b\to s\gamma$ constraint would require
$m_{H^\pm}>500$ GeV \cite{Misiak2} for which $H^\pm\to tb$ would be the dominant decay channel.

The first study of the dependence of the BRs of $H^\pm$ in 3HDMs in terms
of the parameters $\tan\beta, \tan\gamma, \theta,$ and $\delta$ (which determine the values of $X$, $Y$, and $Z$) was given in 
\cite{Akeroyd:2016ssd}, with further detailed studies in \cite{Akeroyd:2018axd}.
It was shown that a large BR$(H^\pm\to cb)$ with $m_{H^\pm}< m_t$ (i.e. the condition $|X|\gg |Y|,|Z|$ is possible) 
can be obtained in the flipped and democratic 3HDMs only.

We now briefly mention other models in which a large BR($H^\pm\to cb$) is possible. In the 2HDM 
(Type III) the fermions receive their masses from both VEVs. Consequently, there are 
scalar FCNCs at tree level, and these are suppressed by small couplings instead of an alignment of Yukawa matrices.
The Yukawa couplings of $H^\pm$ in the 2HDM 
(Type III) depend on more parameters  than in the A2HDM and hence a large BR($H^\pm\to cb$) can be obtained \cite{HernandezSanchez:2012eg}.
Similar comments apply to 3HDMs without NFC \cite{Ivanov:2021pnr} and Four-Higgs-doublet models with NFC \cite{Logan}. In models for which $X, Y,$ and $Z$ 
depend on several parameters one expects some parameter 
space for $|X|\gg |Y|,|Z|$ and thus the possibility of a large BR$(H^\pm\to cb)$ for $m_{H^\pm} < m_t$ while satisfying the $b\to s \gamma$ constraint.
\\

\section{Searches for $t\to H^\pm b$ for at the LHC}
\noindent
Before the commencement of the LHC, searches for $e^+e^-\to H^+H^-$ at LEP2 obtained limits on $m_{H^\pm}$ in the range $74\to 90$ GeV \cite{ALEPH:2013htx}
for the decay channels $H^\pm\to \tau\nu$ and $H^\pm\to cs+cb$ (called "hadronic channel" in which the $s$, $c$ and $b$ quarks are not distinguished), with the assumption BR$(H^\pm\to \tau\nu)$+BR$(H^\pm\to cs+cb)=1$.
The dominant production mechanism at the LHC for an $H^\pm$ being lighter than the top quark is the 
process $pp\to  t\overline t$ followed by the decay $t\to H^\pm b$.
Prior to the LHC, searches in this channel were carried out at the Fermilab Tevatron  \cite{Abazov:2009aa,Aaltonen:2009ke}
 (using $p\overline p\to t\overline t$), but the sensitivity
to BR($t\to H^\pm b$) is much greater at the LHC. 

From the $t\overline t$ pair,  the signal is taken to be one top quark decaying conventionally via $t\to Wb$ (with a BR very close to 1) and the other top quark decays via $t\to H^\pm b$
i.e. the signal is $t\overline t\to H^\pm b W^\pm b$.
The case of both top quarks decaying to $H^\pm b$ gives a negligible number of events. At the LHC four decay channels of $H^\pm$ have been searched for:
$\tau\nu$, $cs+cb$, $cb$, and $A^0W$ with subsequent decay $A^0\to \mu^+\mu^-$. The latter search \cite{CMS:2019idx} requires an on-shell $W$ (and hence $m_A < m_{H^\pm}-m_W$), and will not be considered in this 
work. From the lack of any statistically significant signal, limits are obtained on the products BR($t\to H^\pm b)\times {\rm BR} (H^\pm\to \tau\nu/cs+cb/cb)$, which will be
discussed in detail below. 
Taking $|V_{tb}|=1$ and neglecting small terms that depend on $m_b$ (apart from $m_b$ in Yukawa coupling of $H^\pm$) one has the following expressions for 
the decays of a top quark to a $W$ boson or an $H^\pm$:
\begin{eqnarray}
\Gamma(t\to W^\pm b)=\frac{G_F m_t}{8\sqrt 2 \pi}[m_t^2+2m_W^2][1-m_W^2/m_t^2]^2\,,  \\ \nonumber
\Gamma(t\to H^\pm b)=\frac{G_F m_t}{8\sqrt 2 \pi}[m^2_t|Y|^2 + m_b^2|X|^2][1-m_{H^\pm}^2/m_t^2]^2\,.
\end{eqnarray}
As can be seen from the above equations, BR($t\to H^\pm b$) depends on 
the magnitude of $|X|$ and $|Y|$. As discussed in section II,
the BRs of $H^\pm$ depend on the {\sl relative} values of $|X|$, $|Y|$ and $|Z|$.
 The LHC has accumulated around $139$ fb$^{-1}$
of integrated luminosity at $\sqrt s=13$ TeV. Not all of this data has been used yet in the 
searches for $t\to H^\pm b$, which are summarised in Table~\ref{LHC_search}.

\begin{table}[h]
\begin{center}
\begin{tabular}{|c||c|c|}
\hline
$\sqrt s \;($integrated luminosity)& ATLAS &  CMS  \\ \hline
7 TeV (5 fb$^{-1}$)
&  $cs$ \cite{Aad:2013hla}, $\tau\nu$ \cite{Aad:2012rjx,Aad:2012tj}
&  $\tau\nu$ \cite{Chatrchyan:2012vca} \\
8 TeV (20 fb$^{-1}$)
& $\tau\nu$ \cite{Aad:2014kga} & $cs$ \cite{Khachatryan:2015uua}, 
$cb$ \cite{CMS:2018dzl}, 
$\tau\nu$ \cite{Khachatryan:2015qxa}  \\
13 TeV (36 fb$^{-1}$)
& $cb$\cite{ATLAS:2021zyv}, $\tau\nu$ \cite{ATLAS:2018gfm} & $cs$ \cite{CMS:2020osd}, $\tau\nu$  \cite{CMS:2019bfg}\\
\hline
\end{tabular}
\end{center}
\caption{Searches for $H^\pm$ at the LHC, using $pp\to t\overline t$ and $t\to H^\pm b$. The integrated
luminosities for the searches are given next to the collider energy $\sqrt s$, the exception being the search for $cb$ \cite{ATLAS:2021zyv} at 13 TeV by ATLAS, which used 139 fb$^{-1}$.}.
\label{LHC_search}
\end{table}

\subsection{Decay $H^\pm \to \tau\nu$}
\noindent
For the decay $H^\pm\to \tau\nu$ there are four basic signatures which arise from  $t\overline t\to H^\pm b W^\pm b$. 
Each of $H^\pm$ and $W^\pm$ has a leptonic decay mode ($H^\pm\to \tau\nu\to \ell\nu\nu$, $W^\pm\to \mu^\pm\nu,e^\pm\nu$) and a 
hadronic decay mode ($H^\pm \to \tau\nu\to {\rm hadrons}+{\rm several}\,\nu$, $W^\pm\to q\overline q$).
Only a subset of these 
signatures has been searched for in the two searches below.

A CMS search was carried out with 13 TeV data and 36 fb$^{-1}$ \cite{CMS:2019bfg}.
The following three signatures were searched for:\\
(i) leptonically ($e^\pm,\mu^\pm$) decaying $W^\pm$ and
hadronically decaying $\tau$.\\
(ii) hadronically decaying $W^\pm$ and hadronically decaying $\tau$.\\
(iii) leptonic final state without a hadronically decaying $\tau$. \\
The limits are
obtained by combining these three separate searches.
Significantly improved upper limits on  
BR$(t\to H^\pm b)\times {\rm BR}(H^\pm\to \tau\nu$) 
were obtained, ranging from $< 0.36\%$ for $m_{H^\pm}=80$ GeV to
$<0.08\%$ for $m_{H^\pm}=160$ GeV. The limit for
$m_{H^\pm}=130$ GeV is roughly $< 0.14\%$.

There has been a search with the 13 TeV data \cite{ATLAS:2018gfm}
from the ATLAS collaboration using 36 fb$^{-1}$. Two signatures were targeted, these being the leptonic and hadronic decays of the $W^\pm$ boson
where the $\tau$ is taken to decay hadronically in both cases. No limits are presented 
for the region 80 GeV $\le m_{H^\pm} \le 90$ GeV (unlike the CMS search), but similar limits on 
BR$(t\to H^\pm b)\times {\rm BR}(H^\pm\to \tau\nu$) to the CMS search in \cite{CMS:2019bfg} were obtained in the range
$90\,{\rm GeV} < m_{H^\pm} < 160$ GeV.
The limit for
$m_{H^\pm}=130$ GeV is roughly $< 0.11\%$.

\subsection{Search for $H^\pm \to cs/cb$}
\noindent
For $m_{H^\pm}< m_t$ the dominant hadronic decay modes are $H^\pm\to cs$ and $H^\pm\to cb$. 
Other decay channels to two quarks are suppressed by small quark masses and/or small CKM matrix elements.
LHC searches have been carried out that are sensitive to the sum of BR$(H^\pm\to cs)$ and BR$(H^\pm\to cb)$, which
we will label as BR$(H^\pm\to cs+cb)$. In the publications of these LHC searches,  BR$(H^\pm\to cs)$ is assumed to be 
much larger than BR$(H^\pm\to cb)$, and hence the signal is labelled as $"H^\pm\to cs"$ instead of
"$H^\pm\to cs+cb$". We will use the latter labelling, as we will focus on the case of BR$(H^\pm\to cb)$ being 
comparable or greater in magnitude than BR$(H^\pm\to cs)$.

The first search for $H^\pm\to cs+cb$ at the LHC was by ATLAS \cite{Aad:2013hla} with 5 fb$^{-1}$ of data at 7 TeV.
A search was then carried out by CMS \cite{Khachatryan:2015uua} using 20 fb$^{-1}$ of data at 8 TeV. In \cite{Khachatryan:2015uua} the $W$ boson
is taken to decay leptonically. A $b-$tag requirement is used to identify the two $b-$quarks that arise from the decay of the
$t-$quarks. The presence of $H^\pm$ would show up as a peak at $m_{H^\pm}$ in the invariant mass
distribution of the two quarks that are not $b-$tagged
(which are assumed to be the $c$ and $s$ quarks that
originate from $H^\pm$).
 From the lack of any statistically significant signal, limits
on the product BR$(t\to H^\pm b)\times {\rm BR}(H^\pm\to cs + cb$) are obtained, 
which range from around $< 5\%$ for $m_{H^\pm}= 90$ GeV to  $<2\%$ for  $m_{H^\pm}=160$ GeV.
These limits are weaker than those from $H^\pm\to \tau\nu $ decay
for a given $m_{H^\pm}$. In \cite{Khachatryan:2015uua} there are no limits in the region 80 GeV $\le m_{H^\pm} \le 90$ GeV. This
is because  the dominant background from $W\to qq$ decays gives rise to a peak centred on around 80 GeV. 

A search for $H^\pm\to cs+cb$ with 36 fb$^{-1}$ of data at 13 TeV was carried out by the CMS collaboration in
\cite{CMS:2020osd}. In addition to the increased integrated luminosity and centre-of-mass energy compared to the search
in \cite{Khachatryan:2015uua}, charm tagging on the $c$ quark from $H^\pm \to cs$ decay was used to further increase the
sensitivity. Limits of around $<1\%$ are set on the mass range 80 GeV $\le m_{H^\pm} \le 90$ GeV (a region for which
there was no limit in \cite{Khachatryan:2015uua}), improving to 
around $< 0.3\%$ for 100 GeV $\le m_{H^\pm} \le 160$ GeV. The limit for
$m_{H^\pm}=130$ GeV is $< 0.27\%$.

\subsection{Search for $H^\pm \to cb$}
\noindent
Early phenomenological discussions of a direct search for $H^\pm\to cb$ at high-energy colliders by implementing a $b$-tag (to distinguish this channel from
$H^\pm\to cs$ and to reduce backgrounds from $W\to ud/cs$) can be found in \cite{Grossman,Akeroyd:1994ga,Akeroyd:1998dt} in the context of LEP2 (at which no dedicated search 
for $H^\pm\to cb$ was carried out).
The possibility of $t\to H^\pm b$ followed by $H^\pm\to cb$ at hadron colliders (Tevatron and LHC) was first mentioned in \cite{Akeroyd:1995cf} and later
in \cite{Logan:2010ag} (the latter in the context of the flipped 2HDM). A first rough estimate of the gain in sensitivity that could be achieved by tagging the 
$b$-quark from $H^\pm\to cb$ as well as a study of the parameter space of $|X|$, $|Y|$ and $|Z|$ that could be probed at the LHC in the channel $t\to H^\pm b, \,H^\pm \to cb$
was given in \cite{Akeroyd2}.

Motivated by the possibility of a large BR$(H^\pm\to cb)$ in the models described in section II, two dedicated searches
have been carried out for $H^\pm \to cb$ at the LHC. The search differs from that for $H^\pm\to cs+cb$ due the
extra requirement of a third tagged $b$ quark (from $H^\pm\to cb$), which suppresses any contribution to the signal from $H^\pm\to cs$. 
The CMS search  \cite{CMS:2018dzl} for $H^\pm \to cb$ is with 20 fb$^{-1}$ of data at 8 TeV, and 
uses the leptonic ($e^\pm$, $\mu^\pm$) decay of $W$.
Signal events have three $b-$quarks, 
and a fitting procedure was carried out in order to correctly identify the tagged 
$b-$quark that arises from $H^\pm\to cb$. This $b-$quark is then used (together with the non-$b$-tagged $c$ quark) in the invariant mass distribution of $H^\pm$. 
The extra $b-$tag reduces the backgrounds (e.g. $W\to ud, cs$ in the decay $t\to Wb$) relative to the search for $H^\pm\to cs+cb$.  
Moreover, the background from $W\to cb$, which has a $b$ quark, is very suppressed due to the small value of the CKM matrix element $|V_{cb}|$. The limits on BR$(t\to H^\pm b)\times {\rm BR}(H^\pm\to cb$)
are around $< 1.4\%$ for $m_{H^\pm}=90$ GeV, and strengthen
with increasing $m_{H^\pm}$ to $< 0.3\%$ for $m_{H^\pm}=150$ GeV. No limits are given in the mass range 80 GeV $\le m_{H^\pm}\le 90$ GeV.
These limits are stronger than those for $H^\pm \to cs+cb$ for a given $m_{H^\pm}$ with the same $\sqrt s$ and integrated luminosity
i.e. comparing the limits in  \cite{Khachatryan:2015uua} for $H^\pm \to cs+cb$ with those in \cite{CMS:2018dzl} for $H^\pm \to cb$
(both with 20 fb$^{-1}$ of data at $\sqrt s=8$ TeV) one sees that the expected limits on $H^\pm\to cb$ are roughly a factor of two stronger than those
for  $H^\pm \to {\rm hadrons}$. The limit for $m_{H^\pm}=130$ GeV is approximately $< 0.40\%$.

Recently a search has been carried out by ATLAS \cite{ATLAS:2021zyv} using 139 fb$^{-1}$ of integrated luminosity with $\sqrt s=13$ TeV.
A neural network with thirty input variables is used to separate the $H^\pm\to cb$ signal from the background.
The search in \cite{ATLAS:2021zyv} has an expected sensitivity of BR$(t\to H^\pm b)\times {\rm BR}(H^\pm\to cb)\approx 0.1\%$ in the mass region 60 GeV $\le m_{H^\pm}\le 150$ GeV. This 
is a significant improvement over the sensitivity in  \cite{CMS:2018dzl}, which is $0.6\%\to 0.8\%$ in the range 90 GeV $\le m_{H^\pm}\le 150$ GeV. Moreover, a limit is also
obtained for the region 60 GeV $\le m_{H^\pm}\le 90$ GeV (which is not covered in  \cite{CMS:2018dzl}). The observed limits in \cite{ATLAS:2021zyv} are always above the expected limits
for a given $m_{H^\pm}$.
For 60 GeV $\le m_{H^\pm}\le 110$ GeV the observed limit varies between 0.15\% and 0.20\%. For $m_{H^\pm}=120$ GeV, 130 GeV, 140 GeV and 150 GeV the observed limits are approximately
0.25\%, 0.30\%, 0.25\% and 0.20\% respectively.

\subsection{Local excess of $3\sigma$ at $m_{H^\pm}=130$ GeV in the search for $H^\pm\to cb$ by ATLAS} 
\noindent
In the ATLAS search \cite{ATLAS:2021zyv} there is a local excess of around $3\sigma$ around $m_{H^\pm}=130$ GeV. The global significance is $1.6\sigma$.
So far there has been no CMS search with 139 fb$^{-1}$ of integrated luminosity with $\sqrt s=13$ TeV, and (as mentioned above) the CMS search with $\sqrt s=8$ TeV
gave an observed upper limit of $<0.4\%$ for $m_{H^\pm}=130$ GeV.
Taking this excess as genuine, the best-fit value is BR$(t\to H^\pm b)\times {\rm BR}(H^\pm\to cb)=0.16\%\pm 0.06\%$. In the context of several models with an $H^\pm$ we will show in section IV the 
region of parameter space that provides this best-fit value, while respecting constraints from the searches for $H^\pm\to cs+cb$, $H^\pm\to \tau\nu$, and the measurement of BR$(b\to s\gamma)$.

In our earlier work \cite{Akeroyd2},\cite{Akeroyd:2018axd} we studied the magnitude of BR$(t\to H^\pm b)\times {\rm BR}(H^\pm\to cb$) and 
BR$(t\to H^\pm b)\times {\rm BR}(H^\pm\to cs+cb$) in specific models in which a large value of BR($H^\pm \to cb$) is possible.
In  \cite{Akeroyd2} the dependence of the above products of BRs (for $m_{H^\pm}=80$ GeV and 120 GeV) on $|X|$ and $|Y|$ with $|Z|=0.1$ was 
presented in which $|X|$,$|Y|$,$|Z|$ were taken as independent parameters. Contours of BR$(t\to H^\pm b)\times {\rm BR}(H^\pm\to cb$) were plotted in the
plane $[|X|,|Y|]$ with the value of the smallest contour being 0.2\% i.e. roughly the same as the above best-fit value of $0.16\%\pm 0.06\%$.
In \cite{Akeroyd:2018axd} the work of  \cite{Akeroyd2}  was extended to the case of the lightest $H^\pm$ in 3HDMs with NFC. In  \cite{Akeroyd:2018axd} an updated version of the above plot
in the $[|X|,|Y|]$ plane (with $|Z|=0.1$) was given with the value of the smallest contour now being 0.1\% and $m_{H^\pm}=130$ GeV. However, the main purpose of \cite{Akeroyd:2018axd}  was to
study the dependence of BR$(t\to H^\pm b)\times {\rm BR}(H^\pm\to cb$) and 
BR$(t\to H^\pm b)\times {\rm BR}(H^\pm\to cs+cb$) on $\tan\beta,\tan\gamma, \theta$  and $\delta$.
The magnitude of these products of BRs was plotted in the plane $[\tan\gamma,\tan\beta]$, depicting contours of 0.1\% and 0.5\% (and higher values) 
 with $m_{H^\pm}=80$ GeV and 130 GeV. 
 
 Although the parameter space for BR$(t\to H^\pm b)\times {\rm BR}(H^\pm\to cb)=0.16\%\pm 0.06\%$ can be approximately read off from various plots
 in \cite{Akeroyd2} (in terms of $|X|$,$|Y|$,$|Z|$)  and \cite{Akeroyd:2018axd} (in terms of $\tan\beta,\tan\gamma, \theta,\delta$), in this work
 we present some important updates of those earlier works:\\ 
 (i) we clearly depict in the region BR$(t\to H^\pm b)\times {\rm BR}(H^\pm\to cb)=0.16\%\pm 0.06\%$ in the planes $[|X|,|Y|]$ and $[\tan\gamma,\tan\beta]$.
 \\
 (ii) in the works of \cite{Akeroyd2} and \cite{Akeroyd:2018axd} the value $|Z|=0.1$ was taken, but in this work we will display results for several values of $|Z|$.
 Importantly, we will discuss the maximum allowed value of $|Z|$ that is consistent with the $3\sigma$ excess.\\
 (iii) we impose the upper limits on BR$(t\to H^\pm b)\times {\rm BR}(H^\pm\to \tau\nu$) and BR$(t\to H^\pm b)\times {\rm BR}(H^\pm\to cs+cb$), which will
 restrict the parameter space that is consistent with the  $3\sigma$ excess.

\section{Results}
\noindent
Assuming the $3\sigma$ excess to be genuine and resulting from a $H^\pm$ of mass 130 GeV with $0.10\% \le {\rm} {\rm BR}(t\to H^\pm b)\times {\rm BR}(H^\pm\to cb)\le 0.22\%$,
in this section we study the parameter space in three separate classes of models that could give rise to such a signal: \\
(i) a model with one $H^\pm$ and $|X|$, $|Y|$, $|Z|$ taken as independent parameters.\\
(ii) 2HDMs with NFC. \\
(iii) 3HDMs with NFC. \\
Case (i) is well approximated by the A2HDM. We will also impose the constraint on the parameter space of $H^\pm$ from a lack of signal in the channels 
$t\to H^\pm b$ with $H^\pm\to {\rm hadrons}$ or $H^\pm\to \tau\nu$, as well as the constraint from the measurement of BR($b\to s\gamma$).  

\begin{figure}[phtb]
\includegraphics[width=79mm]{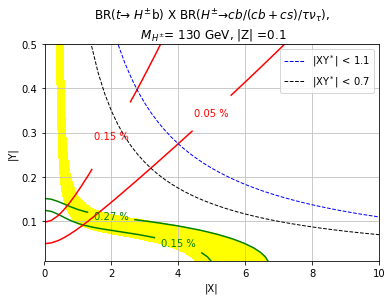}\hspace{4.0mm}
\includegraphics[width=79mm]{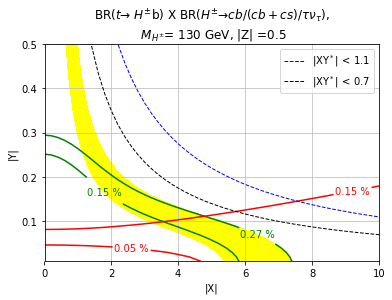}
 \caption{Left panel: The shaded/yellow region in the plane $[|X|,|Y|]$ corresponds to
 BR$(t\to H^\pm b)\times{\rm BR}(H^\pm\to cb)=0.16\%\pm 0.06\%$
with $m_{H^\pm}=130$ GeV and $|Z|=0.1$. Also depicted are  i) contours of BR$(t\to H^\pm b)\times{\rm BR}(H^\pm\to cs+cb)$ with values
0.27\% (current limit) and 0.15\% (future Run II limit); ii) contours of BR$(t\to H^\pm b)\times{\rm BR}(H^\pm\to \tau\nu)$ with values 0.15\%
(current limit) and 0.05\% (future Run II limit); iii) contours of the constraint $b\to s\gamma$ with the upper (lower) curve corresponding to
Re$(XY^*)= -1.1 (0.7)$. The allowed parameter space lies below these contours. Right panel: Same as the left panel, but with $|Z|=0.5$.}
\label{MH130XYplane}
\end{figure}

Fig.~\ref{MH130XYplane} (left panel) is for case (i) in which the region in the plane $[|X|,|Y|]$ that gives
$0.10\% \le {\rm} {\rm BR}(t\to H^\pm b)\times {\rm BR}(H^\pm\to cb)\le 0.22\%$ for $m_{H^\pm}=130$ GeV is displayed.
This is a model independent approach in which $|X|$,$|Y|$ and $|Z|$ are taken as independent
parameters, but this scenario also has an interpretation in the A2HDM.
Three separate figures are shown, each with a different value of $|Z|$  (we take $|Z|=0.1, 0.5$ and $0.9$). The region consistent with measurements of 
BR($b\to s\gamma$) lies 
below the curves of $|XY^*|\le 0.7$ or $|XY^*|\le 1.1$, depending on the sign of 
Re$(XY^*)$ in eq.~(\ref{bsgamma}). 
As discussed in section III, the current upper limit on BR$(t\to H^\pm b)\times {\rm BR}(H^\pm\to \tau\nu$) for $m_{H^\pm}=130$ GeV is
around 0.11\% from ATLAS and 0.14\% from CMS, with both limits using 36 fb$^{-1}$ of integrated luminosity at $\sqrt s=13$ TeV. On Fig.~(\ref{MH130XYplane}) we (conservatively) take this upper limit to be 0.15\%, and
we also show a contour of 0.05\% which might be attainable with the full Run II integrated luminosity of 139 fb$^{-1}$.
Also depicted is the upper limit on BR$(t\to H^\pm b)\times {\rm BR}(H^\pm\to cs+cb$) for $m_{H^\pm}=130$ GeV, which is 0.27\% from CMS with 36 fb$^{-1}$ of integrated luminosity at $\sqrt s=13$ TeV
(no Run II search yet from ATLAS).
We also show a contour of 0.15\% which might be attainable with the full Run II integrated luminosity of 139 fb$^{-1}$ in this channel.
In Fig.~\ref{MH130XYplane} (left panel) the shaded/yellow region in each plot corresponds to contours of BR$(t\to H^\pm b)\times {\rm BR}(H^\pm\to cb$) of 0.22\% (upper) and 0.10\% (lower).
For the case of $|Z|=0.1$, all of the shaded/yellow region lies below the $b\to s\gamma$ contours, and much of the region survives the current constraints
from $H^\pm\to cs+cb$ and $H^\pm\to \tau\nu$ i.e. a sizeable part of a rectangle defined by $0.5 < |X|<6.5$ and $|Y|<0.15$ would
give the required product of BRs. If the excess is genuine, then a signal in the channels $H^\pm\to \tau\nu$ and $H^\pm\to cs+cb$ would also start to show (with 139 fb$^{-1}$)
in those parts of the shaded/yellow region that lie between the contours 0.15\% and 0.05\% for ($H^\pm\to \tau\nu$) and between the contours 0.27\% and 0.15\% ($H^\pm\to cs+cb$).
There is a small shaded/yellow region that lies below the 0.05\% and  0.15\% contours, for which no signal for $H^\pm\to \tau\nu$ and $H^\pm\to cs+cb$ would start to show
with the Run II data (139  fb$^{-1}$). In the shaded/yellow region for a fixed value of $|Y|$ the value of ${\rm BR}(H^\pm\to cb)$ will be larger as $|X|$ increases. 
In Table~\ref{points} for specific values
of  $|X|$, $|Y|$ and $|Z|$ the values of BR$(t\to H^\pm b)$, BR$(H^\pm\to cb)$ and BR$(t\to H^\pm b)\times$BR$(H^\pm\to cb)$ are given in this region (for $m_{H^\pm}=130$ GeV).

\begin{table}[h]
\begin{center}
\begin{tabular}{|c|c|c|c|c|c|c|}
\hline
$m_{H^\pm}$ & $|X|$ & $|Y|$ & $|Z|$ &  BR$(t\to H^\pm b)$ &  BR$(H^\pm\to cb)$  &  BR$(t\to H^\pm b)\times$BR$(H^\pm\to cb)$  \\ \hline
130 GeV & 1  &  0.10 & 0.10 & 0.232\%& 46.5\% & 0.108\% \\ \hline
130 GeV & 2 & 0.10 & 0.10 & 0.250\% &  68.7\% & 0.172\% \\ \hline
130 GeV & 4 & 0.05 & 0.10 & 0.154\% &  78.9\% & 0.121\% \\ \hline
130 GeV & 4 &  0.10 & 0.50 &0.322\% & 44.9\% &  0.145\% \\ \hline
130 GeV & 6 & 0.05 & 0.50 & 0.275\% & 60.1\% & 0.165\% \\ \hline
130 GeV & 7 & 0.01 & 0.50 & 0.299\% & 64.7\% &  0.193\% \\ \hline
\end{tabular}
\end{center}
\caption{Values of BR$(t\to H^\pm b)$, BR$(H^\pm\to cb)$ and BR$(t\to H^\pm b)\times$BR$(H^\pm\to cb)$ for specific values
of  $|X|$, $|Y|$ and $|Z|$ which give rise to  $0.10\% \le {\rm} {\rm BR}(t\to H^\pm b)\times {\rm BR}(H^\pm\to cb)\le 0.22\%$ (for $m_{H^\pm}=130$ GeV). }.
\label{points}
\end{table}

\begin{figure}[phtb]
\includegraphics[width=79mm]{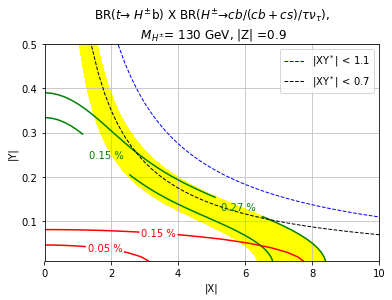}
 \caption{Contours of BR$(t\to H^\pm b)\times{\rm BR}(H^\pm\to cb)$
in the plane $[|X|,|Y|]$ with $m_{H^\pm}=130$ GeV and $|Z|=0.9$.}
\label{MH130XYplane09}
\end{figure}

In Fig.~\ref{MH130XYplane} (right panel) we take  $|Z|=0.5$, which increases BR$(H^\pm\to \tau\nu)$ relative to Fig.~\ref{MH130XYplane} (left panel). The shaded/yellow region now shifts to the right because larger values of $|X|$ are needed
to maintain $0.10\% \le {\rm} {\rm BR}(t\to H^\pm b)\times {\rm BR}(H^\pm\to cb)\le 0.22\%$ with the larger value of $|Z|$. A sizeable part of a rectangle defined by $4 < |X|<7$ and $|Y|<0.15$
would explain the excess, and would guarantee a signal in the channel $H^\pm\to \tau \nu$ with all the Run II data because the entire shaded/yellow region lies above
the 0.05\% contour. No signal in the channel $H^\pm\to \tau \nu$ with 139  fb$^{-1}$ would disfavour the interpretation of
the excess being genuine for this value of $|Z|=0.5$. In Fig.~\ref{MH130XYplane09} we take $|Z|=0.9$, which further increases BR($H^\pm\to \tau\nu)$. One can see that only a small part (around $|X|=7$) of the shaded/yellow region
lies below the 0.15\% contour, and if the excess is genuine a signal would start to show in the $H^\pm\to \tau\nu$ channel with all the Run II data. For values of $|Z|>1$, the yellow/shaded region moves further to the right, and 
it is not possible to simultaneously respect the current limit on $H^\pm\to \tau\nu$ (0.15\% contour), $b\to s\gamma$, and have $0.10\% \le {\rm} {\rm BR}(t\to H^\pm b)\times {\rm BR}(H^\pm\to cb)\le 0.22\%$ for $m_{H^\pm}=130$ GeV. For values $|Z|>1$, the contour 0.05\% for the $H^\pm\to \tau\nu$ search rules out all of the $[X,Y]$ plane except for a region of small $|X|$ and $|Y|$.


\begin{figure}[phtb]
\includegraphics[width=79mm]{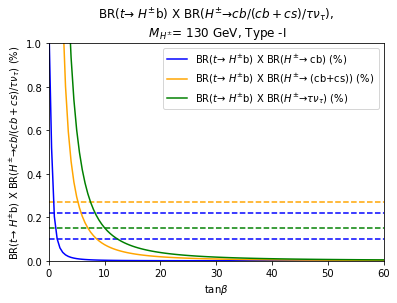}\hspace{4.0mm}
\includegraphics[width=79mm]{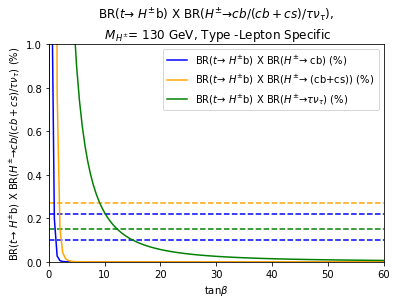}
 \caption{Left panel: In the 2HDM (Type I) with $m_{H^\pm}=130$ GeV, on the $y-$axis the three products BR$(t\to H^\pm b)\times{\rm BR}(H^\pm\to cb)$, BR$(t\to H^\pm b)\times{\rm BR}(H^\pm\to cs+cb)$ and
 BR$(t\to H^\pm b)\times{\rm BR}(H^\pm\to \tau\nu)$ are displayed (by solid lines) as a function of $\tan\beta$. Also depicted are the current upper limits (dotted lines) on the channels $\tau\nu$ ($<0.15\%)$ and $cs+cb$ 
 ($<0.27\%$), and
 BR$(t\to H^\pm b)\times{\rm BR}(H^\pm\to cb)=0.16\%\pm 0.06\%$.
 Right panel: Same as the left panel, but for the 2HDM (Lepton-specific).}
\label{MH130type13}
\end{figure}

\begin{figure}[phtb]
\includegraphics[width=79mm]{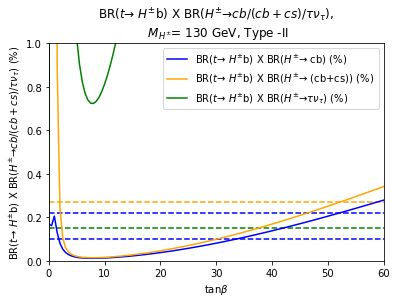}\hspace{4.0mm}
\includegraphics[width=79mm]{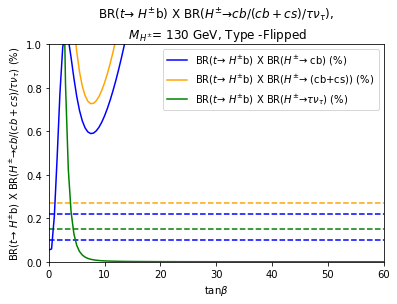}
\caption{Left panel: In the 2HDM (Type II) with $m_{H^\pm}=130$ GeV, on the $y-$axis the three products BR$(t\to H^\pm b)\times{\rm BR}(H^\pm\to cb)$, BR$(t\to H^\pm b)\times{\rm BR}(H^\pm\to cs+cb)$ and
 BR$(t\to H^\pm b)\times{\rm BR}(H^\pm\to \tau\nu)$ are displayed (by solid lines) as a function of $\tan\beta$. Also depicted are the upper limits (dotted lines) on the channels $\tau\nu$ 
 ($<0.15\%)$  and $cs+cb$ , and ($<0.27\%$), and 
 BR$(t\to H^\pm b)\times{\rm BR}(H^\pm\to cb)=0.16\%\pm 0.06\%$.
 Right panel: Same as the left panel, but for the 2HDM (Flipped).}
\label{MH130type24}
\end{figure}

Having discussed the excess in the context of a 2HDM with independent couplings $|X|$, $|Y|$ and $|Z|$ we now turn our attention to 2HDMs with NFC, of which there are four
distinct types (as discussed in Section IIA).
In such models these couplings are not independent and depend on just one parameter $\tan\beta$, as shown in Table~I. 
In Fig.~\ref{MH130type13} the $y-$axis refers to any of the three products BR$(t\to H^\pm b)\times{\rm BR}(H^\pm\to cb)$, BR$(t\to H^\pm b)\times{\rm BR}(H^\pm\to cs+cb)$ and
 BR$(t\to H^\pm b)\times{\rm BR}(H^\pm\to \tau\nu)$, which are displayed by solid lines as a function of $\tan\beta$ ($x-$axis).  The dotted horizontal lines depict the
 upper bounds for the searches for $H^\pm\to \tau\nu$ and $H^\pm\to {\rm hadrons}$, as well as the region BR$(t\to H^\pm b)\times{\rm BR}(H^\pm\to cb)=0.16\%\pm 0.06\%$.
The left panel is for the 
 2HDM (Type I) and the right panel is for the 2HDM (Type II), with $m_{H^\pm}=130$ GeV.
 Fig.~\ref{MH130type24} is the same as Fig.~\ref{MH130type13}  but for the 2HDM (Lepton-specific) in the left panel and the 2HDM (Flipped) in the right panel.
 As discussed in Section IIB, the decay $b\to s\gamma$ constrains $X$, $Y$ for a particular value of $m_{H^\pm}$. In the 2HDM (Type II) and 2HDM (Flipped) one has
 $XY^*=\tan\beta\cot\beta=1$, leading to $m_{H^\pm}> 500$ GeV \cite{Ciuchini1,Ciuchini2,Borzumati,Gambino,Misiak,Misiak2}.
 Hence $m_{H^\pm}=130$ GeV is not possible in either of these models. However,  if an extension of the SM has either of these 2HDM structures as well as 
 additional particles that also contribute to $b\to s\gamma$ (e.g. 
 the Minimal Supersymmetric SM (MSSM), which has the form of a 2HDM (Type II) but with charginos $\chi^\pm_1$ and $\chi^\pm_2$) then
 $m_{H^\pm}=130$ GeV might be possible due to destructive interference between $H^\pm$ and the additional particles in the prediction for
 $b\to s\gamma$.  Hence we show results in both the 2HDM (Type II) and 2HDM (Flipped), assuming that additional particles in the model can
 allow $m_{H^\pm}=130$ GeV to be compatible with $b\to s\gamma$.
  Note that in the MSSM with R-parity conservation there is no coupling between $\chi^\pm_i$ and two fermions, and thus 
 $\chi^\pm_i$ cannot give a signal identical to $t\to H^\pm b$.
 In the 2HDM (Type I) and 2HDM (Lepton-specifc) one has $XY^*=-\cot^2\beta$. The bound on Re$(XY^*)$ in eq.~(\ref{bsgamma})
 (which roughly applies to $m_{H^\pm}=130$ GeV) can be satisfied for appropriately chosen values of $\cot\beta$ (i.e. $\cot^2\beta< 1.1$, giving approximately the bound $\tan\beta >1$).
 
 In Fig.~\ref{MH130type13} (left panel) for the 2HDM (Type I) it can be seen that there is only a very small region around $\tan\beta=1$ that predicts 
  BR$(t\to H^\pm b)\times{\rm BR}(H^\pm\to cb)=0.16\%\pm 0.06\%$, but this region of $\tan\beta$ is ruled out from the searches for $H^\pm\to cs+cb$ and
  $H^\pm\to \tau\nu$, which require approximately $\tan\beta > 5$ and $>10$ respectively. A similar behaviour is seen in Fig.~\ref{MH130type13} (right panel) for the 2HDM (Lepton-specific), for which
  $H^\pm\to cs+cb$ and $H^\pm\to \tau\nu$ require approximately $\tan\beta > 2$ and $>12$ respectively, while BR$(t\to H^\pm b)\times{\rm BR}(H^\pm\to cb)=0.16\%\pm 0.06\%$ is
  again obtained only for $\tan\beta\approx 1$. As mentioned earlier, on these plots the $b\to s\gamma$ constraint is roughly given by $\tan\beta >1$. 
  Hence neither of these models can explain the $3\sigma$ excess, but both allow an $H^\pm$ of 130 GeV for $\tan\beta > 10$ (in Type I for which the BRs of $H^\pm$ are
  independent of $\tan\beta$) and $\tan\beta > 12$ (in Lepton-specific, for which BR($H^\pm\to \tau\nu$) would dominate for $\tan\beta>12$). More generally, in both
  of these models there always exists a parameter space of larger values of $\tan\beta$ for which the possibility of the decay $t\to H^\pm b$ is not ruled out.

In Fig.~\ref{MH130type24} (left panel) for the 2HDM (Type II) it can be seen that there are two regions that predict 
  BR$(t\to H^\pm b)\times{\rm BR}(H^\pm\to cb)=0.16\%\pm 0.06\%$, these being roughly $1 < \tan\beta < 2$ and  $33 < \tan\beta < 52$. However, the search for
   $H^\pm\to \tau\nu$ rules out all values of $\tan\beta$ and thus the excess at $m_{H^\pm}=130$ GeV cannot be accommodated in the 2HDM (Type II).
   In Fig.~\ref{MH130type24} (right panel) for the 2HDM (Flipped) it can be seen that there is only one region that predicts
  BR$(t\to H^\pm b)\times{\rm BR}(H^\pm\to cb)=0.16\%\pm 0.06\%$, this being roughly $\tan\beta\approx 1$. However, the search for
   $H^\pm\to \tau\nu$ rules out $\tan\beta<5$, while the search for $H^\pm\to cs+cb$ rules out all values of $\tan\beta$. As discussed in
    Section IIC, the 2HDM (Flipped) is the only 2HDM with NFC that has a parameter space for a large BR$(H^\pm\to cb)$. This 
    can be seen on the plots, which show an increasingly large value for BR$(t\to H^\pm b)\times{\rm BR}(H^\pm\to cb)$ as $\tan\beta$ increases. 
   In conclusion, none of the four 2HDMs with NFC can accommodate the excess at $m_{H^\pm}=130$ GeV, and only the Type I and Lepton-specific models
  allow a parameter space of $\tan\beta$ for which $m_{H^\pm}< m_t-m_b$.  

\begin{figure}[phtb]
\includegraphics[width=79mm]{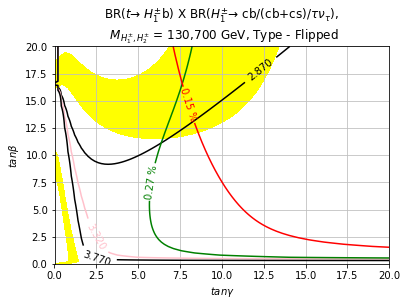}\hspace{4.0mm}
\includegraphics[width=79mm]{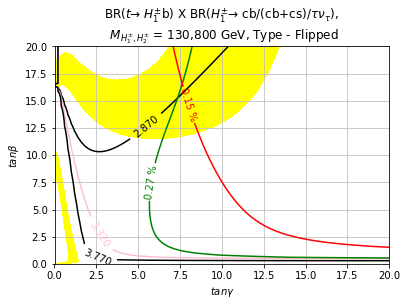}
 \caption{Left panel: In the 3HDM (Flipped) with $m_{H^\pm_1}=130$ GeV and $m_{H^\pm_2}=700$ GeV,
 the shaded/yellow region in the plane $[\tan\gamma,\tan\beta]$ corresponds to
 BR$(t\to H^\pm_1 b)\times{\rm BR}(H^\pm_1\to cb)=0.16\%\pm 0.06\%$. Also depicted are  i) contours of BR$(t\to H_1^\pm b)\times{\rm BR}(H^\pm_1\to cs+cb)=0.27\%$ and
BR$(t\to H^\pm_1 b)\times{\rm BR}(H^\pm_1\to \tau\nu)=0.15\%$ (current experimental upper limits); ii) contours of BR$(b\to s\gamma)$ with the upper ($3.77\times 10^{-4}$)
and lower ($2.87\times 10^{-4}$) limits at $3\sigma$, and the experimental central value of $3.32\times 10^{-4}$. The allowed parameter space lies between the contours of 
3.77 and 2.87, and to the right of the contours of 0.27\% and 0.15\%.   
 Right panel: Same as the left panel, but for $m_{H^\pm_2}=800$ GeV.}
\label{flip_700}
\end{figure}

The previous plots were for 2HDMs in different scenarios. We now turn our attention to a third class of models i.e. 3HDMs with NFC. As discussed in section II, in 3HDMs 
there are two charged scalars (labelled by $H^\pm_i$ with $i=1,2$) and the couplings $X_i$, $Y_i$ and $Z_i$
depend on the four parameters $\tan\beta$, $\tan\gamma$, $\theta$ and $\delta$.
Of the five distinct 3HDMs listed in Table~\ref{valuesudl} we focus on the 3HDM (Flipped) for which BR$(H^\pm\to cb)$ can be dominant.
Fig.\ref{flip_700} is similar to  Fig.~\ref{MH130XYplane} and Fig.~\ref{MH130XYplane09} but in the plane $[\tan\gamma,\tan\beta]$ for the 3HDM (Flipped) 
with $m_{H^\pm_1}=130$ GeV and $m_{H^\pm_2}=700$ (left panel) and $m_{H^\pm_2}=800$ GeV (right panel). We take $\theta=-\pi/2.1$ and $\delta=0$.
 As before, the shaded/yellow region in the plane $[\tan\gamma,\tan\beta]$) corresponds to
 BR$(t\to H^\pm_1 b)\times{\rm BR}(H^\pm_1\to cb)=0.16\%\pm 0.06\%$. Also depicted are  i) contours of BR$(t\to H_1^\pm b)\times{\rm BR}(H^\pm_1\to cs+cb)=0.27\%$ and
BR$(t\to H^\pm_1 b)\times{\rm BR}(H^\pm_1\to \tau\nu)=0.15\%$ (current experimental upper limits); ii) contours of BR$(b\to s\gamma)$ with the upper ($3.77\times 10^{-4}$)
and lower ($2.87\times 10^{-4}$) limits at $3\sigma$, and the experimental central value of $3.32\times 10^{-4}$. The calculation of BR$(b\to s\gamma)$ is done
with the contributions of $H^\pm_1$ and  $H^\pm_2$
at Next-to-Leading Order, using the results from our previous work \cite{Akeroyd:2020nfj}.

In Fig.\ref{flip_700} the allowed parameter space lies between the contours of 
3.77 and 2.87 (the constraint from BR$(b\to s\gamma))$, and to the right of the contours of 0.27\% and 0.15\% (the constraint from LHC searches for $H^\pm_1\to cs+cb$ and $H^\pm_1\to \tau\nu$ respectively).
In the left panel of Fig.\ref{flip_700} (with $m_{H^\pm_2}=700$ GeV), it can be seen that most of the shaded/yellow region of BR$(t\to H^\pm_1 b)\times{\rm BR}(H^\pm_1\to cb)=0.16\%\pm 0.06\%$ is ruled out, but
there is small region inside a rectangle given by $12.5 < \tan\beta < 20$ and $7.5 < \tan\gamma < 12.5$ that satisfies the above constraints. Despite this large mass splitting between the two 
charged scalars, we expect that the contribution of these scalars to the $S$, $T$, and $U$ parameters \cite{Peskin:1990zt} can be kept within the experimental limits. This is due to the large number of neutral Higgs bosons in the 
3HDM (three CP-even and two CP-odd)
which would also be present in the one-loop corrections to electroweak precision observables from the charged scalars, and allow the possibility of cancellation among any sizeable contributions.
A specific study of $S$, $T$ and $U$ in a 3HDM has been carried out in \cite{Kalinowski:2021lvw}.

\begin{figure}[phtb]
\includegraphics[width=79mm]{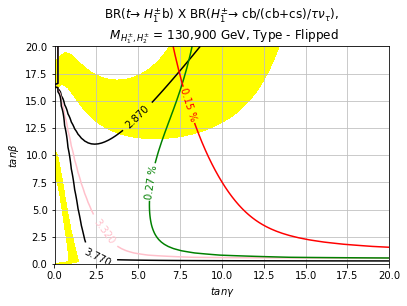}\hspace{4.0mm}
\includegraphics[width=79mm]{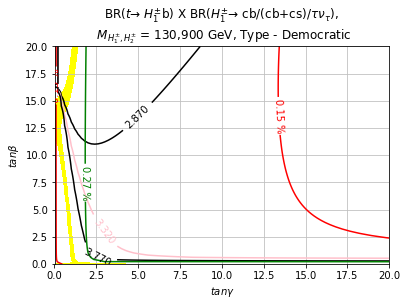}
\caption{Left panel: Same as Fig.\ref{flip_700} but for for $m_{H^\pm_2}=900$ GeV. Right panel: Same as the left panel, but for the 3HDM (Democratic).}
\label{flip_900}
\end{figure}

As $m_{H^\pm_2}$ is decreased below 
700 GeV this (small) allowed region shrinks, and then vanishes.
 In the right panel of Fig.\ref{flip_700}
(with $m_{H^\pm_2}=800$ GeV) one can see that the contours for BR$(b\to s\gamma$) shift with respect to their location for $m_{H^\pm_2}=800$ GeV (all other contours do not move
as these only depend on the value of $m_{H^\pm_1}$). The contour for $2.87\times 10^{-4}$ (which crosses the shaded/yellow region), moves to the left and thus a larger region of
BR$(t\to H^\pm_1 b)\times{\rm BR}(H^\pm_1\to cb)=0.16\%\pm 0.06\%$ in the 
rectangle $12.5 < \tan\beta < 20$ and $7.5 < \tan\gamma < 12.5$ now satisfies all constraints. In the left panel of Fig.\ref{flip_900} we take $m_{H^\pm_2}=900$ GeV, and 
the allowed region of BR$(t\to H^\pm_1 b)\times{\rm BR}(H^\pm_1\to cb)=0.16\%\pm 0.06\%$ further increases in size. In the right panel of Fig.\ref{flip_900}
we take the same input parameters as for the left panel, but in the 3HDM (Democratic). One can see that the yellow/shaded region has decreased substantially in size (due to the change of model) but is ruled out
(it lies to the left of the contours of 0.27\% and 0.15\%). 

In all of the above plots for the 3HDM the CP-violating phase $\delta$ is taken to be zero, and our results show that $m_{H^\pm_2}$ should be heavy ($> 700$ GeV)
if $H^\pm_1$ is to accommodate the excess at 130 GeV. Taking $\delta>0$ allows both $H^\pm_1$ and $H^\pm_2$ to be lighter than the top quark (as shown in \cite{Akeroyd:2016ssd,Akeroyd:2020nfj})
while respecting the constraint from $b\to s\gamma$. However, the couplings $X_i$, $Y_i$ and $Z_i$ would then have an imaginary part. This leads to a non-zero value for the EDM of the neutron, for which there is a stringent upper limit. 
 As shown in \cite{Logan:2020mdz}, a very restricted parameter space with $\delta>0$ and $m_{H^\pm_{1,2}}< m_t$ can simultaneously satisfy the constraints from $b\to s\gamma$ and the EDM of the neutron. However,
 with the additional constraint on $X_1$, $Y_1$ and $Z_1$ from accommodating the excess at 130 GeV it is difficult to find parameter space to satisfy all constraints simultaneously. If such a parameter space
 is found then  $H^\pm_2$ could also be produced in the decay of the top quark via $t\to H^\pm_2 b$. As mentioned above, the  couplings $X_2$, $Y_2$ and $Z_2$ (which also depend on $\tan\beta$, $\tan\gamma$, $\theta$, 
 and $\delta$)  
 are constrained by the requirement of $X_1$, $Y_1$ and $Z_1$ having values that explain the excess at 130 GeV. This will be explored elsewhere.

We comment that $H^\pm$ in the above models with couplings that accommodate the excess at 130 GeV would not give
a sizeable contribution to the ratios of leptonic $B$ meson decays $R(D)$ and $R(D^*)$, for which the experimental measurements are somewhat above the SM
predictions (see e.g. \cite{London:2021lfn}). This is because the $H^\pm$ contribution to $R(D)$ and $R(D^*)$ depends on the product $|XZ|$, but 
the parameter space that accommodates the excess at 130 GeV has moderate ($<10$) values of $|X|$ and small values of $|Z|(<1)$. Larger values of
$|XZ|$ would be needed to enhance $R(D)$ and $R(D^*)$ sufficiently.

If this excess at 130 GeV turns out to be genuine, such an $H^\pm$ could be studied in detail at a future $e^+e^-$ collider via the process $e^+e^-\to H^+H^-$, provided that
$\sqrt s > 2m_{H^\pm}>260$ GeV.
The decays $H^\pm\to \tau\nu$ and $H^\pm\to cs$ can be measured precisely if their BRs are of the order of a few percent or more.
The prospects for precision measurements of $H^\pm \to cb$ at future $e^+e^-$ colliders have been discussed in
\cite{Akeroyd:2019mvt}, with a recent detailed simulation in \cite{Hou:2021qff}.

\section{Conclusions}  
\noindent
Searches for $H^\pm$ in the channel $t\to H^\pm b$ at the LHC now include the decay channel $H^\pm\to cb$, which can be the dominant decay mode for $H^\pm$ in regions of parameter space of specific models with
two or more scalar doublets \cite{Grossman,Akeroyd:1994ga}. The first search was by CMS in 2018 \cite{CMS:2018dzl} (at $\sqrt s=8$ TeV with 20 fb$^{-1}$) 
and the second search was in 2021 by ATLAS  \cite{ATLAS:2021zyv} (at $\sqrt s=13$ TeV with 139 fb$^{-1}$).
A local excess of around $3\sigma$ (global $1.6\sigma$) has been observed in the search by ATLAS and the excess is best fitted by $m_{H^\pm}$ of around 130 GeV 
and a product of BRs given by BR$(t\to H^\pm b)\times{\rm BR}(H^\pm\to cb)=0.16\%\pm 0.06\%$. 
Treating this slight excess as genuine and building on our previous work in this search channel \cite{Akeroyd2,Akeroyd:2016ssd,Akeroyd:2018axd},
we present the parameter space for which this excess can be accommodated in the context of three classes of models with two or more scalar doublets.
The limits from LHC searches for $t\to H^\pm b$ with subsequent decay
$H^\pm\to cs$ and $H^\pm\to \tau\nu$ at $m_{H^\pm}$=130 GeV are taken into account, as well as the constraint from $b\to s\gamma$.

In the context of 2HDMs with independent $|X|$,$|Y|$ and $|Z|$ couplings for $H^\pm$ (an example being the A2HDM) it is shown that
the excess can accommodated for moderate values of the coupling $|X|$ ($2< |X| <10$), small values of $|Y|$ ($<0.1$) and small values of $|Z|$$(<1)$,
giving $40\% < {\rm BR}(H^\pm \to cb) < 80\%$.

It was then shown that such an excess cannot be explained in 2HDMs with natural flavour conservation.
In the flipped 3HDM with no extra sources of CP-violation in the $H^\pm$ couplings ($\delta=0$)
the excess can be accommodated by $H^\pm_1$ in a restricted parameter space of $\tan\beta,\tan\gamma$ and $\theta$,  provided that $m_{H_2^\pm} >700$ GeV.
Forthcoming searches with 139 fb$^{-1}$ at $\sqrt s=13$ TeV in the channels $H^\pm\to cb$ (CMS), $H^\pm \to cs$ (ATLAS/CMS) and $H^\pm \to \tau\nu$ (ATLAS/CMS)
should clarify whether the excess is the first sign of an $H^\pm$ with a mass of around 130 GeV.

\section*{Acknowledgements}
SM is funded in part through the NExT Institute and the STFC CG ST/L000296/1.

\end{document}